\documentclass[english,twocolumn,aps]{revtex4}
\usepackage{amssymb}
\usepackage{graphicx}
\usepackage{color}
\usepackage{pgf}

\usepackage{tikz}
\usetikzlibrary{arrows}
\usetikzlibrary{trees,snakes}
\tikzstyle{block}=[draw opacity=0.7,line width=1.4cm]

\definecolor{red}{rgb}{1,0,0}
\definecolor{green}{rgb}{0,1,0}
\definecolor{blue}{rgb}{0,0,1}

\def\be{\begin{equation}}
\def\ee{\end{equation}}
\def\bsp{\be\begin{split}}

\def\a{\alpha}
\def\b{\beta}
\def\g{\gamma}

\def\e{\epsilon}
\def\m{\mu}

\def\s{\sigma}

\begin{document}

\title{Antiparticles as particles: QED$_{2+1}$ and composite Fermions  at $\nu = \frac{1}{2}$}
\author{Abhishek Agarwal}
\affiliation{American Physical Society, Ridge, NY, 11961, \\
email: abhishek@aps.org}

\begin{abstract}
We present a continuum operator map that inverts the role of particles and antiparticles in three dimensional QED. This is accomplished by the attachment of specific holomorphic Wilson lines to Dirac fermions. We show that this nonlocal map provides a continuum realization of Son's composite fermions at $\nu = \frac{1}{2}$. The inversion of Landau levels and the Dirac-cone--composite-Fermion duality is explicitly demonstrated for the case of slowly varying magnetic fields. The role of Maxwell-terms as well as the connection of this construction to a gauge-invariant formulation of 3D gauge theories is also elaborated upon.
\end{abstract}

\maketitle
\section{Introduction}Given $QED$ in $D = 2+1$, can one find a dual theory such that the antiparticles of $QED$ are realized as particles in the dual gauge theory? In this paper we present a construction of such a duality while working within a Hamiltonian framework in the continuum. Our search for a particle-hole symmetric description of QED is motivated by  the recently proposed duality\cite{son1} between the Dirac cone action
\be
S_1 = -i\int \bar \Psi \gamma ^\mu (\partial_\mu - iA_\mu) \Psi +\cdots \label{t1}
\ee
and the composite fermion theory defined by the action
\be
S_2 =  -i\int \bar \chi \gamma ^\mu (\partial_\mu - ia_\mu)\chi -\frac{1}{4\pi}\int \epsilon^{\mu \nu\rho}a_\mu \partial_\nu A_\rho +\cdots \label{t2}
\ee
which is expected to provide a particle-hole symmetric description of quantum hall states \cite{son1,son2,son3}.
In the second theory $A$ is to be regarded as an external electromagnetic field while $a$ is dynamical.  (In our convention $\{\g^\m = i\s^3, \s^1, \s^2\}$.) \\ A key property of (\ref{t2}) is that the composite fermions do not couple directly to the photon through the usual covariant derivative\cite{son1}, but do so  via their density $\rho_\chi = \langle \chi^\dagger \chi\rangle = \frac{B}{4\pi}$. $B$ is magnetic field associated with $A$. On the other hand the electromagnetic change density $\rho_e = \frac{\delta S_2}{\delta A_0} = -\frac{b}{4\pi}$, where $b$ is the magnetic field associated with $a$. The associated filling fractions $\nu_e = 2\pi \frac{\rho_2}{B}$ and $\nu_\chi = 2\pi \frac{\rho_\chi}{b}$ are related as
\be
2\nu_e = -\frac{1}{2\nu_\chi} \label{fill}
\ee
This inversion is at the heart of the realization of the particle-hole symmetry in the composite fermion picture\cite{son2, son3}. \\
From a quantum field theory point of view, the proposed duality presents a Fermionic version of the bosonic particle-vortex duality and it has been absorbed in the web of 3d dualities that have been the subject of much recent investigation\cite{dual1,dual2,dual3,dual4}. Given the growing importance of this proposed duality in both high energy as well as condensed matter physics, it is imperative to ask if an explicit map between the two theories described by (\ref{t1}) and (\ref{t2}) may be established.
In this paper we present a continuum realization of this duality at $\nu = \frac{1}{2}$ by finding an operator map between QED and a dual gauge theory that inverts the role of particles and antiparticles. Before presenting our construction, we briefly review some salient lessons learned from various approaches to understanding the duality which will play an important role in what is to follow.
\\
Insights from topological insulators (TI) provide important clues about how this duality might work\cite{TI0, TI1}. A key realization in that context is that the IR physics obtained by integrating out $\Psi$ and $\chi$ contains half-quantized Chern-Simons (CS) terms with opposite level numbers\cite{TI1}.The leading IR effective actions are
\be
S_1 \rightarrow S_{1eff} = \pm\frac{1}{8\pi} \int \epsilon^{\mu \nu\rho}A_\mu \partial_\nu A_\rho \label{eff1}
\ee
while
\be
S_2 \rightarrow S_{2eff} = \mp\frac{1}{8\pi} \int \epsilon^{\mu \nu\rho}a_\mu \partial_\nu a_\rho \mp\frac{1}{4\pi}\int \epsilon^{\mu \nu\rho}a_\mu \partial_\nu A_\rho \label{eff2}
\ee
The absolute signs (as we shall later see) in front of the two actions depends on various microscopic details. The relative signs between the various terms in the two actions are what are crucial for the purposes of this paper. For latter convenience, we note that the $a_0$ equation of motion from the second equation above  leads to 
\be
b = -B \label{bB}
\ee
More generally, it is easily seen that eliminating the dynamical $a$ fields from (\ref{eff2}) leads to (\ref{eff1}). While insightful,TI based constructions do not clarify if the duality may be expected to hold independently of the context of TIs, and if it does, does it hold beyond the lowest order in the IR captured by the CS theories above?\\
Lattice approaches - aka coupled wire constructions - are a very promising microscopic approach.  An explicit lattice realization of this duality, using  the formidable machinery of 2d bosonization was recently claimed in\cite{lat} (see alo \cite{lat1,lat2, lattice}). However, as with many lattice constructions, the existence of the continuum limit remains an open issue. The construction presented here is defined directly in the continuum, though it interestingly shares several parallels with \cite{lat}. \\ 
The analysis of the present paper may be taken as complementary to the path integral approaches \cite{dual1,dual2} that focus on relating  currents and partition functions of (\ref{t1}, \ref{t2}). The Hamiltonian approach, combined with the complex structure inherent to the spatial geometry has the advantage of allowing a precise map between microscopic degrees of freedom of QED$_{2+1}$ and its particle-hole dual. 
\section{Gauge fields and the dual photon}
Our starting point is the realization that spatial components of gauge fields $A = \frac{1}{2}(A_1 + iA_2)$ and $\bar A = \frac{1}{2}(A_1 - iA_2)$ can be parameterized in terms of the $SL(N, \mathbb{C})$-valued complex matrices $M$ and $M^\dagger$ as
\be
A = -\partial MM^{-1}, \hspace{.3cm} \bar A = M^{\dagger -1}\bar \partial M^{\dagger}\label{M}
\ee
It is understood that $\partial = \frac{1}{2}(\partial_i + i\partial_2)$ and $z = x_i -ix_2$
This parameterization - which is at the heart of the gauge invariant formulation of gluodynamics initiated in \cite{KKN1} - applies in general to nonabelian $SU(N)$ gauge fields $A = -iA^at^a$, where $t^a$ are the $SU(N)$ generators. The remarkable advantage of this parameterization is that local (time-independent) gauge transformations are simply realized as $M \rightarrow UM$, where $U$ us a unitary matrix. This implies that $H = M^\dagger M$ is gauge invariant and it is the basic physical degree of freedom in terms of which gluodynamics was formulated in \cite{KKN1}. The matrix $M$ can be regarded as a $complex$ Wilson line as it satisfies (from its defining relation) $DM = 0$, where $D = \partial +A$. The real and imaginary parts of this matrix correspond to the physical and gauge degrees of freedoms encoded in $A$ respectively. This is seen most explicitly in the Abelian limit - which is the primary focus of this paper - where we can express
\be
M = e^\theta \label{mab}
\ee
where $\theta$ is a complex scalar. 
In this limit  the parameterization (\ref{M}) reduces to the usual gradient-curl (Hodge) decomposition $A_i =  \partial_i Im(\theta) + \epsilon_{ij}\partial_j Re(\theta)$. One can also define a set of auxiliary gauge fields which we suggestively call $a, \bar a$ - for reasons that would soon be manifest - through the parameterization:
\be
a =  M^{\dagger -1}\partial M^{\dagger}, \hspace{.3cm} \bar a = -\bar \partial MM^{-1} \label{Ma}
\ee
(\ref{M}) and (\ref{Ma}) yield the relations
\be
D\bar a = \bar \partial A, \hspace{.3cm} \bar D a =  \partial \bar A \label{defa}
\ee
which, in the Abelian limit imply (\ref{bB}). We will now focus on the Lowest Landau Levels (LLLs) for fermions coupled to $A$ and $a$ respectively and show that for a given value of $B$, the roles of particles and holes are interchanged for Fermions coupled to $A$ and $a$ respectively. We first consider the case of a constant background magnetic field $B >0$. It is also instructive - especially for comparison with\cite{lat} - to include a mass term for the fermions. The (mass-added) Hamiltonian following from (\ref{t1}) is
\be
H[\Psi] = i\int \bar \Psi (\g^iD_i - m)\Psi = \frac{1}{2} \int \Psi^*\left(\begin{array}{cc}
m & -2iP_-\\
2iP_+ & -m
\end{array}\right)\Psi
\ee
We can  choose the gauge where $A_i = -\frac{1}{2}\epsilon_{ij}x_j B$. $\e_{12} = +1$.  
\be
P_+ = -i\partial -\frac{i}{4} B\bar z, \hspace{.3cm}
P_- = -i\bar\partial + \frac{i}{4}Bz
\ee
The momenta are mapped to the standard oscillator variables as $P_+ = \sqrt{\frac{B}{2}} A$,  $P_- = \sqrt{\frac{B}{2}} A^\dagger$. The spectrum is given by $E^2  - 4P_-P_+ = m^2$ and it is easy to see that $\Psi_0 = \left(\begin{array}{c}\psi_1 \\ 0\end{array}\right): \hspace{.2cm} P_+\psi_1 = 0$ 
is a normalizable zero mode for both $E = \pm m$.  The vacuum is taken to be the state where all the negative energy levels are filled. When $E = +m$,  the lowest energy state has positive energy, so the vacuum $|\Omega_\Psi\rangle$ can be identified with the Fock vacuum $|0\rangle_\Psi$. However,  when $E = -|m|$, the zero-mode must  be regarded as part of the negative energy states. Thus the vacuum $|\Omega_\Psi\rangle$, in this case  is not the Fock vacuum, but must be taken to be 
\be
|\Omega_\Psi \rangle= \alpha^\dagger |0\rangle_\Psi; 
\ee
It is understood that  $\alpha ^\dagger$ is the oscillator for the zero mode in the harmonic expansion of the field given by
\begin{eqnarray}
\Psi = \alpha \Psi_0 + \sum_n \alpha_n\Psi^+_n + \sum_n\beta ^\dagger_n\Psi^-_n
\end{eqnarray}
 where the quantization condition is $\alpha |0\rangle_\Psi = 0$ is implied. 
Using the operator expression
\be
\frac{1}{2}[\Psi^\dagger, \Psi] = \sum_n(\a^\dagger_n\a_n - \b^\dagger_n \b_n) + \frac{1}{2}(\a^\dagger \a - \a \a^\dagger) 
\ee
along with the fact that the degeneracy is given by $\frac{B}{2\pi}$, we can compute the electronic charge density to be
\begin{eqnarray}
\rho_e  &=& \langle \Omega_\Psi | \frac{1}{2}[\Psi^\dagger, \Psi]|\Omega_\Psi \rangle=  \frac{B}{2\pi}\langle \Omega_\Psi |\frac{1}{2} (\a^\dagger \a - \a \a^\dagger) |\Omega_\Psi\rangle \nonumber\\
&&= -\frac{m}{|m|}\frac{B}{4\pi}
\end{eqnarray}
We can also read off the effective action (assuming slowly varying $B$) via
\be
\rho_e =   \frac{\delta}{\delta A_0} S^{Eff}_{\Psi} = -\frac{m}{|m|}\frac{1}{8\pi} \frac{\delta}{\delta A_0}\int \epsilon^{\mu \nu \rho}A_\mu \partial_\nu A_\rho \label{csA}
\ee
The analysis so far recapitulates the well known picture for the emergence of an effective Chern-Simons action from LLL physics. The filling fraction here is $\nu_e = -\frac{1}{2}\frac{m}{|m|}$. However, when one couples fermions $\chi$ (with the $same$ mass term) to the $a$ fields in the same background magnetic field $B = -b$, the resultant Dirac equation is:
\be
\left(\begin{array}{cc}
m & -2iP(a)_-\\
2iP(a)_+ & -m
\end{array}\right) \left(\begin{array}{c}\chi_1 \\ \chi_2\end{array}\right) = E\left(\begin{array}{c}\chi_1 \\ \chi_2\end{array}\right)\label{ev}
\ee
$-2iP(a)_- = -(2\bar \partial + \frac{1}{2}zB)$ and $2iP(a)_+ = (2\partial - \frac{1}{2}\bar{z}B)$. Because of the change in the sign of the magnetic field for $a$ relative to $B$, $P(a)_+$ now acts as the creation operator as opposed to the previous case. The normalizable zero-modes now correspond to $E = \mp m$ for $\frac{m}{|m|} = \pm1$, which is the converse of case of fermions coupled to $A$. Using a mode expansion for $\chi$
\be
\chi = \mathcal{A} \chi_0 + \sum_n \mathcal{A}_n\chi^+_n + \sum_n\mathcal{B} ^\dagger_n\chi^-_n
\ee 
we see that $|\Omega_\chi\rangle$ for $E = -(+)m$ is given by $\mathcal{A}^\dagger |0\rangle_\chi ( |0\rangle_\chi)$.  $\mathcal{A}$ is the oscillator for the zero mode of $\chi$ and $\ |0\rangle_\chi: \mathcal{A}|0\rangle_\chi = 0$. The complete momentum space mode expansion for $\Psi$ and $\chi$ in a constant magnetic field is given in the appendix using which it can be explicitly checked that the Landau levels of $\chi$ are obtained by inverting those for $\Psi$ for a given fixed value of $B$ and $m$. This is depicted in the figure below.\\
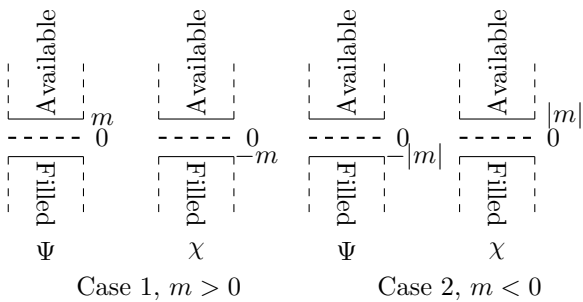
\begin{tikzpicture}
\draw [dashed](0,0.25) -- (0,1);
\draw [dashed](1,0.25) -- (1,1);
\draw (0,0.25) -- (1,0.25);
\node[rotate=90] at (.5,1)  {Available};
\node[rotate=-90] at (.5,-.75)  {Filled};
\draw [thick,dashed](0,0) -- (1,0);
\node at (1.25,0) [,]{$0$};
\node at (1.25,.25) [,]{$m$};
\draw (0,-0.25) -- (1,-0.25);
\draw [dashed](0,-0.25) -- (0,-1);
\draw [dashed](1,-0.25) -- (1,-1);
\draw [thick,dashed](2,0) -- (3,0);
\draw [dashed](2,0.25) -- (2,1);
\draw [dashed](3,0.25) -- (3,1);
\draw (2,0.25) -- (3,0.25);
\draw (2,-0.25) -- (3,-0.25);
\draw [dashed](3,-0.25) -- (3,-1);
\draw [dashed](2,-0.25) -- (2,-1);
\node at (3.25,0) [,]{$0$};
\node at (3.3,-.25) [,]{$-m$};
\node at (.5,-1.5) [,]{$\Psi$};
\node at (2.5,-1.5) [,]{$\chi$};
\node[rotate=90] at (2.5,1)  {Available};
\node[rotate=-90] at (2.5,-.75)  {Filled};
\draw [thick,dashed](4,0) -- (5,0);
\draw [thick,dashed](6,0) -- (7,0);
\draw (4,0.25) -- (5,0.25);
\draw (6,0.25) -- (7,0.25);
\draw (4,-0.25) -- (5,-0.25);
\draw (6,-0.25) -- (7,-0.25);
\draw [dashed](4,0.25) -- (4,1);
\draw [dashed](5,0.25) -- (5,1);
\draw [dashed](6,0.25) -- (6,1);
\draw [dashed](7,0.25) -- (7,1);
\draw [dashed](4,-0.25) -- (4,-1);
\draw [dashed](5,-0.25) -- (5,-1);
\draw [dashed](6,-0.25) -- (6,-1);
\draw [dashed](7,-0.25) -- (7,-1);
\node[rotate=90] at (4.5,1)  {Available};
\node[rotate=90] at (6.5,1)  {Available};
\node[rotate=-90] at (4.5,-.75)  {Filled};
\node[rotate=-90] at (6.5,-.75)  {Filled};
\node at (5.25,0) [,]{$0$};
\node at (7.25,0) [,]{$0$};
\node at (5.4,-.25) [,]{$-|m|$};
\node at (7.4,+.3) [,]{$|m|$};
\node at (4.5,-1.5) [,]{$\Psi$};
\node at (6.5,-1.5) [,]{$\chi$};
\node at (2,-2) [,]{Case 1, $m>0$};
\node at (6,-2) [,]{Case 2, $m<0$};
\end{tikzpicture}
\\
This inversion suggests that the holes of the $\chi$ field depict particles of the $\Psi$ field and vice versa (we will substantiate this further later). Noting that the magnetic field experienced by dual $\chi$ particles is $b$, with the degeneracy given by $\frac{b}{2\pi}$, the induced charge density for $\chi$ is computed to be 
\begin{eqnarray}
\rho_\chi &=&  \langle \Omega_\chi| \frac{1}{2}[\chi^\dagger, \chi]|\Omega_\chi \rangle =+ \frac{m}{|m|} \frac{b}{4\pi} \nonumber\\
&&= \frac{m}{|m|}\frac{\delta}{\delta a_0}\int \left(\frac{1}{8\pi}  \epsilon^{\mu \nu \rho}a_\mu\partial_\nu a_\rho \right)\label{csa} 
\end{eqnarray}
We note that the sign of the Chern-Simons term is reversed compared to (\ref{csA}), even though the sign of the mass-term has $not$ changed. But what action does (\ref{ev}) follow from? The relationship $b=-B$ as well as the first equality in (\ref{csa}) allows us to work backward and see that the action whose quantization in a constant magnetic field background generates (\ref{ev}) is given precisely by
\be
S(\chi) =  -i\int \bar \chi( \gamma ^\mu (\partial_\mu - ia_\mu) -m)\chi +\frac{m}{|m|}\frac{1}{4\pi}\int \epsilon^{\mu \nu\rho}a_\mu \partial_\nu A_\rho \label{sm}
\ee
The parameterization (\ref{Ma}) of the spatial components of $a_\mu$ simply serve to solve the Gauss' law constraint (the $a_0$ equation of motion following from the action above) through the relation $b = -B$.
Using (\ref{csa}), we  also see that $\nu_\chi = +\frac{1}{2}\frac{m}{|m|}$ and satisfies (\ref{fill}). The complete effective action for $\chi$ is given by the Chern-Simons term computed from $\rho_\chi$ as well as the mixed term already present in $S(\chi)$. 
\be
S^{Eff}_{\chi} = \frac{m}{|m|}\frac{1}{8\pi} \int \epsilon^{\mu \nu\rho}a_\mu \partial_\nu a_\rho +\frac{m}{|m|}\frac{1}{4\pi}\int \epsilon^{\mu \nu\rho}a_\mu \partial_\nu A_\rho \label{seffc}
\ee
which agrees with (\ref{eff2}). The two signs in (\ref{eff2}) correspond to the two possible values of $\frac{m}{|m|}$.
\section{Fermions and dual fermions}
We have so far seen that fermions coupled to the auxiliary gauge fields $a$ satisfy the properties of Son's composite fermions at half-filling. However to really establish a duality we need a map between the $\Psi$ and $\chi$ fermions. This is what we do next.
\\Let us define composite fermions $\chi$ in terms of the original spinor $\Psi$ by the following operator map:
\be
 \left(\begin{array}{c}\psi_1\\ \psi_2 \end{array}\right)  = \left(\begin{array}{c}
M (M^{\dagger  -1})^t\chi_2^*\\ M^{\dagger  -1}M^t\chi_1^*\end{array}\right)  =\exp{2iIm(\theta)}\left(\begin{array}{c}
\chi_2^*\\ \chi_1^*\end{array}\right) \label{map1}
\ee
and
\be
 \left(\begin{array}{c}\psi^*_1\\ \psi^*_2 \end{array}\right)  = \left(\begin{array}{c}
(M^{\dagger })^t M^{-1}\chi_2\\ (M^{-1 })^tM^\dagger\chi_1\end{array}\right) = \exp{-2iIm(\theta)}\left(\begin{array}{c}
\chi_2\\ \chi_1\end{array}\right) \label{map2}
\ee
A straightforward computation now shows that under this map,
\be
H[\Psi] = i\int \bar \Psi (\g^iD_i  - m)\Psi = i\int \bar \chi (\g^id_i - m)\chi = H[\chi]
\ee
where $d_i$ is the standard covariant derivative in which $A$ is replaced by $a$. 
This gives us an explicit operator map from the original electronic theory to the Hamiltonian for the dual ($\chi$) fermions directly in the continuum. This is the central result of the paper which we  explore further below. \\
A few words about the nature of the dual fermions are in order at this point.
First, we note  that (\ref{map1}, \ref{map2}) preserve the symplectic structure; i.e. the dual fermions obey canonical anticommutation relations (this is only true for the Abelian theory).  Further, the mass terms for the two fermions are mapped to each other without a relative sign being introduced, just as was reported in the lattice construction\cite{lat}. The mapping of the mass terms to each other is also crucial for consistency with the LLL analysis performed earlier. In the abelian case, we can explicitly express $M$ as a holomorphic Wilson line. $M = \exp \int _{-\infty}^z iA$.  The factor $\exp{2iIm(\theta)}$ in the definition of $\chi$ can thus be viewed as a field dependent gauge transformation generated by attaching two complex Wilson lines to the original fermion $\Psi$.  The factors of $M$ in (\ref{map1}, \ref{map2}) thus play a role that is very reminiscent of the monopole operators used in \cite{dual1,dual2}. The Hamiltonian approach used in this paper allows us to  generate explicit expressions for them via the complex parameterization (\ref{M}, \ref{Ma}). Further, we note that both $(\chi  \Psi )\rightarrow e^{iIm(\theta)} (\chi  \Psi )$ under $U(1)$ gauge transformations. The common $U(1)$ charge allows us to interpret $\chi$ as as describing bonafide particles (as opposed to simply antiparticles of $\Psi$) coupled to the gauge field $a_\mu$. Finally, it is instructive to cross-check the maps (\ref{map1}, \ref{map2}) with the mode expansions for the spinors $\Psi$ and $\chi$ in constant magnetic field backgrounds, which corresponds to a gauge where $Im(\theta) = 0$. Using the explicit mode expansions presented in the Appendix, we see that (\ref{map1}, \ref{map2}) are satisfied in the Abelian limit as operator maps if we identify 
\be
\alpha = \mathcal{A}^\dagger, \mathcal{B}_n = \alpha _n \hspace{.1cm} \mbox{and} \hspace{.1cm} \mathcal{A}_n = \beta_n. \label{operators} 
\ee
As an added consistency check we note that:
\be
\langle \Omega_\chi| \frac{1}{2}[\chi^\dagger, \chi]|\Omega_\chi \rangle = \langle \Omega_\Psi \frac{1}{2}[\chi^\dagger, \chi]|\Omega_\Psi \rangle
\ee
The left hand side refers to the computation already presented; where $|\Omega_\chi \rangle$ was computed after explicitly imposing the constraint $B = -b$ on the $a$ field and the quantization condition for $|0\rangle _\chi$ was the usual one: $\mathcal{A}|0\rangle _\chi = 0$. The right hand side uses (\ref{operators}) and makes use of the fact that the quantization condition for $\chi$ changes on $|\Omega_\Psi\rangle$, since (\ref{operators}) implies that $\mathcal{A}^\dagger |0\rangle_\psi = 0$. The $BF$ term in (\ref{seffc}) imposes precisely this constraint/altered quantization condition. This is further evident when we note that integrating out $A_\mu$ in (\ref{seffc}) amounts to averaging over the constraints and it yields $ S^{Eff}_\chi = -\frac{m}{|m|}\frac{1}{8\pi} \int \epsilon^{\mu \nu\rho}a_\mu \partial_\nu a_\rho$; which is the expected Chern-Simons term for $\chi$ coupled to gauge fields $a_\mu$ without any constraints vis-a-vis $A_\mu$.\\
The analysis presented above extends to the massless case. The parameter $m$ can be regarded as a regulator that makes it manifest that the spectra of $H[\Psi]$ and $H[\chi]$ approach the zero energy state (in the massless limit) from opposite directions (this is clear from the figure in the previous page).  Once the sign of $m$ is fixed, it can be smoothly be taken to zero while ensuring that the previous analysis continues to hold in the massless limit.
\section{Maxwell terms}Finally we focus on making sense of the ellipses in (\ref{t1}, \ref{t2}), which stand for potential Maxwell terms. Having analyzed the mapping of the Fermionic hamiltonians, we now show how the Maxwell terms for the $A$ and $a$ fields transform into each other under (\ref{map1}, \ref{map2}). For this purpose we employ the gauge invariant Hamiltonian framework for 3d gauge fields coupled to fermions developed in \cite{AA-VPN-2}, where it was shown how the photonic Hamiltonian is sensitive to the choice of gauge invariant Fermionic variables. In particular, the sign of the induced Chern-Simons term uniquely dictates how gauge invariant Fermionic variables may be constructed by attaching holomorphic Wilson lines to Dirac Fermions. For the purposes of specificity, we fix $\frac{m}{|m|} = -1$. Following \cite{AA-VPN-2}, the gauge invariant Fermionic variable compatible with the induced CS term for $a$ is
\be
\tilde \Lambda = \left(\begin{array}{c}\tilde \lambda_1\\ \tilde \lambda_2 \end{array}\right)  = \left(\begin{array}{c}
\mathcal{M}^{\dagger }\chi_1\\ \mathcal{M}^{-1 } \chi_2\end{array}\right) \label{covi2} 
\ee
$\mathcal{M} = M^{\dagger -1}$. This change of variables can be regarded as a 2d chiral transformation, whose Jacobian can be expressed as a Wess-Zumino-Witten (WZW) functional \cite{AA-VPN-2}. The measures on the fermionic Hilbert spaces transform under the above as \cite{AA-VPN-2}
\be
d\Psi^* d\Psi \rightarrow  e^{-2 \mathcal{S}(H)} d\check{\tilde{\Lambda}} d\tilde{\Lambda}, d\chi^*d\chi \rightarrow  e^{-2 \mathcal{S}(H^{-1})} d\check{\tilde{\Lambda}} d\tilde{\Lambda}
\ee
$\check{\tilde{\Lambda}}$ stands for the canonical conjugate of $\tilde{\Lambda}$. 
 $\mathcal{S}$ denotes the WZW functional\cite{AA-VPN-1, AA-VPN-2} and it is the boundary piece of the Chern-Simons term induced by the Fermions in the effective action picture.  $\mathcal{S}$ induces a mass term (the topological mass) for in the photonic Hamiltonian. Defining the currents.
 \be
J_l = \partial H H^{-1}\label{J},  J_r = -H^{-1}\partial H\label{j}
\ee
the Maxwell Hamiltonians for the $A$ and $a$ fields take on the following forms.
\begin{eqnarray}
H[A] &=& \frac{e^2}{2\pi}\int \left[-J_l\frac{\delta}{\delta J_l} + \frac{1}{(x-y)^2}\frac{\delta}{\delta J_l(x)}\frac{\delta}{\delta J_l(y)}\right]  \nonumber\\
&&+\frac{1}{2e^2}\int B^2  
\end{eqnarray}
and 
\begin{eqnarray}
H[a] &=& \frac{e^2}{2\pi}\int \left[- J_r\frac{\delta}{\delta  J_r} + \frac{1}{(x-y)^2}\frac{\delta}{\delta  J_r(x)}\frac{\delta}{\delta  J_r(y)}\right] \nonumber\\
&&+ \frac{1}{2e^2}\int b^2  
\end{eqnarray}
We will refer to \cite{AA-VPN-1, AA-VPN-2} for a derivation of the gauge invariant  photonic Hamiltonians for a given choice of gauge invariant fermionic degrees of freedom (we present a brief outline in the Appendix). For the purposes of this discussion we draw attention to the fact that  the first term in $H[A]$ and $H[a]$  acts as  mass terms for the photon, and they are a direct consequence of a the nontrivial measure on the configuration space. The nonlocal factor in the second term in the Hamiltonian is nothing but the two-point function in the OPE of the WZW model\cite{KKN1}. Noting that $J = \partial (Re (\theta))$,  $\tilde J = -\partial (Re (\theta))$ and $b=-B$, it is trivial to see that the two Hamiltonians map to each other term by term.
As with the fermionic Hamiltonian analyzed earlier, the photonic analysis continues to hold when $m = 0$ so long as care is taken to let $m$ approach zero from a fixed side of the real axis. Finally, we point out that the WZW functional is well known in the context of 2d bosonization which, in turn, was central to the lattice approach to this duality\cite{lat}. The interchange of the left and right currents $J_l \leftrightarrow J_r$ between the $A$ and $a$ descriptions is very reminiscent of a similar phenomenon observed on the lattice\cite{lat}. It would be extremely interesting to explore if a closer connection exists between the gauge invariant continuum approach presented here and lattice bosonization methods.
\section{Conclusions and future directions} In the present paper we have shown that d= 2+1 QED with a single species of Dirac Fermion can be mapped to a dual description via an operator map (\ref{map1}) and (\ref{map2}) where:\\
I: The dual fermions possess properties expected of Son's composite fermions including the inversion of filling fractions (\ref{fill}). The operator map explicitly inverts the roles of particles and holes while endowing both $\Psi$ and $\chi$ with the same $U(1)$ transformation properties.\\ 
II: The gauge field $A$ is mapped to a dual photon $a$ defined by (\ref{defa}) which enforces the constraint $B = -b$ for general non-constant magnetic fields. The relationship between the magnetic fields can be regarded as  solution to the Gauss' law constraint  for the action of the composite fermion(\ref{sm}).\\
III: The effective actions (\ref{csA}, \ref{seffc}) obtained by integrating out the fermions match with expectations from TI based constructions.(\ref{map1}) and (\ref{map2}) allow us to understand the mixed BF term in the effective action as enforcing altered quantization conditions required by the operator map.\\
IV: Parity violating mass terms for the fermions  - which map to each other -  acts as  regulators in this analysis. The entire analysis continues to hold hold in the $m=0$ limit so long as care is taken to let $m$ approach zero keeping its sign fixed. The upper and lower signs in (\ref{eff1},\ref{eff2}) correspond to $m$ approaching zero from the positive and negative halves of the real axis respectively. \\
Finally, a gauge invariant formulation, allows us to map the corresponding Maxwell terms of the two theories to each other in the $A_0 = a_0 = 0$ gauges.\\
While not a hindrance to the analysis (including the formal operator maps) presented in this paper, the theories studied in this paper are known to be anomalous (as seen  from the half-quantized induced Chern-Simons term in the effective actions). How one cancels the anomaly will depend on the context in which this duality is employed. A potential remedy could be to add a second Fermioinic species to both sides of the duality with a large parity violating mass term. Taking the mass of the second fermion to infinity will render it non-dynamical while restoring gauge invariance of the partition function.\\ Several additional issues also require further study. The constraint $B=-b$ restricts the construction to $\nu = \frac{1}{2}$. Can this analysis be generalized to other filling fractions? How much of the present analysis can one generalize to the nonabelian case? Following the tell-tale signs  noted earlier, is there  a possible connection between the gauge-invariant framework described here and  bosonization techniques employed in the coupled-wire constructions? We hope future work will clarify these issues.\\
{\bf Acklowledgments:} The author is deeply grateful to Parameswaran Nair for numerous conversations about this work.


\section{Appendix}
\subsection{Mode Expansion}
The complete mode expansions for $\Psi$ and $\chi$ in constant magnetic field backgrounds can be written as:
\begin{widetext}
\begin{eqnarray}
\Psi &=& \alpha  \left(\begin{array}{c}N e^{-Bz\bar{z}/4} \\ 0\end{array}\right) + \nonumber \\
&&\sum_{n\geq 1} \alpha_n \left(\begin{array}{c}M_nz^n \\M_{n-1}\frac{1}{\sqrt{2Bn}}(\sqrt{m^2 +2Bn} -m)z^{n-1} \end{array}\right) e^{-Bz\bar{z}/4}\nonumber \\
&&\sum_{n\geq 1} \beta ^\dagger _n \left(\begin{array}{c}-M_nz^n \\\tilde{M}_{n-1}\frac{1}{\sqrt{2Bn}}(\sqrt{m^2 +2Bn} +m)z^{n-1} \end{array}\right) e^{-Bz\bar{z}/4}\nonumber\\
\end{eqnarray}
\begin{eqnarray}
\chi &=& \mathcal{A}  \sigma_1\left(\begin{array}{c}N e^{-Bz\bar{z}/4} \\ 0\end{array}\right)^* + \nonumber \\
&&\sum_{n\geq 1} \mathcal{A}_n \sigma_1\left(\begin{array}{c}-M_nz^n \\\tilde{M}_{n-1}\frac{1}{\sqrt{2Bn}}(\sqrt{m^2 +2Bn} +m)z^{n-1} \end{array}\right)^* e^{-Bz\bar{z}/4}\nonumber\\
&&\sum_{n\geq 1} \mathcal{B}^\dagger_n \sigma_1\left(\begin{array}{c}M_nz^n \\M_{n-1}\frac{1}{\sqrt{2Bn}}(\sqrt{m^2 +2Bn} -m)z^{n-1} \end{array}\right)^* e^{-Bz\bar{z}/4}\nonumber \\
\end{eqnarray}
\end{widetext}
$N, M_i, \tilde{M}_i$ are normalization constants.
\subsection{Discrete Symmetries}
The dual  fields $a$ and $\chi$ inherit the discrete symmetry transformation properties of $A$ and $\Psi$. The charge conjugation transformations are:
\begin{eqnarray}
CA_\mu C^{-1}  &=& -A_\mu\nonumber\\
Ca_\mu C^{-1}  &=& -a_\mu\nonumber\\
C\Psi C^{-1}  &=& \sigma_1 \Psi^* \mu\nonumber\\
C\chi C^{-1}  &=& \sigma_1 \chi^* \mu\nonumber\\
\end{eqnarray}
Parity transsformation $x_2 \rightarrow -x_2$ acts on the fields as:
\begin{eqnarray}
PA_2 (t,x_1,x_2) P^{-1}  &=& -A_2 (t,x_1,-x_2) \nonumber\\
Pa_2 (t,x_1,x_2) P^{-1}  &=& -a_2 (t,x_1,-x_2) \nonumber\\
P\Psi (t,x_1,x_2) P^{-1}  &=& \gamma^2\Psi (t,x_1,-x_2) \nonumber\\
P\chi (t,x_1,x_2) P^{-1}  &=& \gamma^2\chi(t,x_1,-x_2) \nonumber\\
\end{eqnarray}
The time reversal transformations are given by
\begin{eqnarray}
TA_0 (t,x_1,x_2) T^{-1}  &=& A_0 (-t,x_1,x_2) \nonumber\\
TA_i (t,x_1,x_2) T^{-1}  &=& -A_i (-t,x_1,x_2) \nonumber\\
Ta_0 (t,x_1,x_2) T^{-1}  &=& a_0 (-t,x_1,x_2) \nonumber\\
Ta_i (t,x_1,x_2) T^{-1}  &=& -a_i (-t,x_1,x_2) \nonumber\\
T\Psi (t,x_1,x_2) T^{-1}  &=& -\gamma^1\Psi (-t,x_1,x_2) \nonumber\\
T\chi (t,x_1,x_2) T^{-1}  &=& -\gamma^1\chi (-t,x_1,x_2) \nonumber\\
\end{eqnarray}
The discrete transformation properties listed above do not match with the discrete poperties of the dual gauge fields and fermions listed in \cite{son1}. However, we can define three discrete symmetries $\mathcal{C} = C, \mathcal{P} = CP, \mathcal{T} = CT$. It is readily checked that $\mathcal{C,P,T}$ satisfy the properties of the operators assigned as charge conjugation, parity and time reversal operators in \cite{son1}.
\subsection{Gauge Invariant Photonic Hamiltonians}
Here we gather some salient results needed to obtain the gauge invariant forms of the Maxwell Hamiltonians.
This issue has been explored in depth for non-Abelian Yang-Mills theories has been explored in detail in\cite{KKN1}. 
The general expression for a nonabelian Yang-Mills Hamiltonian in the presence of fermions in the gauge invariant framework is given by \cite{AA-VPN-2}:
\be
H_{YM} =+\frac{e^2}{2}\int H^{ab}(x)\left[{\cal{G}}p^a\right]^\dagger(x){\cal{G}}p^b(x)\ + \frac{1}{2e^2}\int B^a B^a \label{MaxA}
\ee
$p$ and $\bar p$ are the right and left translation operators on $M$ and $M^\dagger$\cite{KKN1}, acting as $[p^a(y), M(x)] = M(x)(-it^a)\delta^2(x-y)$ and $[\bar{p}^a(y), M^\dagger(x)] = -it^aM^\dagger(y)\delta^2(x-y)$. ${\cal{G}}p(x)$ is a shorthand for $\int_u {\cal{G}}(x,u)p(u)$ and ${\bar{\cal{G}}}\bar p(x) = \int_u {\bar{\cal{G}}}(x,u)\bar p(u)$. $\cal{G}$, $\bar{\cal{G}}$, given by\cite{KKN1}  
\begin{eqnarray}
{\cal{G}}(x,y) = G(x,y)\left[ 1 - e^{-|x-y|^2/\e}H^{-1}(y,\bar x)H(y,\bar y)\right]\nonumber \\
{\bar{\cal{G}}}(x,y)  = \bar{G}(x,y)\left[ 1 - e^{-|x-y|^2/\e}H(x,\bar y)H^{-1}(y,\bar y)\right]
\end{eqnarray}
are the point-split-regularized version of the two-dimensional Green's functions $\bar{G}(x,y) = \frac{1}{\pi(x-y)}$ and  $G(x,y) = \frac{1}{\pi(\bar x-\bar y)}$. It is crucial to note that $\left[{\cal{G}}p^a\right]^\dagger$ refers to the adjoint computed with respect to the measure including the WZW term induced by the fermionic change of variables\cite{AA-VPN-2}. Matters simplify considerably in the ablelian limit where
\be
p = -i\frac{\partial}{\partial \theta}, \hspace{.2cm} \bar{p} = -i\frac{\partial}{\partial \bar{\theta}}
\ee
Fixing $\frac{m}{|m|} = -1$, we follow \cite{AA-VPN-2} where it was shown that the gauuge invariant fermionic variable that induces the WZW functional compatible with $S^{Eff}_{\Psi} = -\frac{m}{|m|}\frac{e^2}{8\pi}\int \epsilon^{\mu \nu \rho}A_\mu \partial_\nu A_\rho$ is given by  
 $\Lambda$:
\be
\Lambda = \left(\begin{array}{c}\lambda_1\\ \lambda_2 \end{array}\right)  = \left(\begin{array}{c}
M^{-1}\psi_1\\ M^{\dagger } \psi_2\end{array}\right)   = (I - Im(\theta)  - Re(\theta) \gamma^5 + \cdots)\Psi \label{covi1}
\ee
$\gamma^5 = \sigma^3$ and the above equation shows that transforming to gauge invariant variables may be regarding as a 2d chiral transformation. 
The measure on the fermionic configuration space transforms under the above change of variables as
\be
d\Psi^* d\Psi =  e^{+2\mathcal{S}(H)} d\check{\Lambda} d\Lambda
\ee
$\check{\Lambda}$ denotes the canonical conjugate to $\Lambda$ and $\mathcal{S}$ denotes the Wess-Zumino-Witten (WZW) functional\cite{AA-VPN-1, AA-VPN-2}. $\mathcal{S}(H)$ is the boundary piece of the Chern-Simons term induced by the Fermions in the effective action picture. As argued in \cite{AA-VPN-2}, the subtle compatibility condition between the induced WZW term and the effective action is that the coefficient of $\mathcal{S}(H)$  be twice the level number of the induced Chern-Simons term\cite{AA-VPN-2}. This compatibility was crucial establishing the gaplessless of the spectrum of various supersymmetric 3d theories within the gauge invariant framework \cite{AA-VPN-1}. 
The physical gauge invariant configuration space is  spanned  by functionals of  $H (= M^\dagger M), \Lambda, \check\Lambda $. The complete Maxwell Hamiltonian acting on this gauge invariant configuration space takes the following form
\begin{eqnarray}
H_{Max} &=& \frac{e^2}{2\pi}\int \left[J_l\frac{\delta}{\delta J_l} + \frac{1}{(x-y)^2}\frac{\delta}{\delta J_l(x)}\frac{\delta}{\delta J_l(y)}\right] \nonumber\\
&&+ \frac{1}{2e^2}\int B^2  
\end{eqnarray}
A lengthly and somewhat subtle computation allows one to obtain the above from (\ref{MaxA}) by rewriting the action of (\ref{MaxA}) on functionals of $H (= M^\dagger M), \Lambda, \check\Lambda $ term by term. These are detailed in\cite{KKN1, AA-VPN-1, AA-VPN-2}.\\ 
As far as the dual fermions are concerned, denoting the complex matrix parameterizing $a$ by $\mathcal{M} = M^{\dagger -1}$ (so we may write $a = -\partial \mathcal{M} \mathcal{M}^{-1}$), the reversed sign of its induced Chern-Simons term implies \cite{AA-VPN-2} that the invariant variables corresponding to $\chi$ should be chosen as:
\be
\tilde \Lambda = \left(\begin{array}{c}\tilde \lambda_1\\ \tilde \lambda_2 \end{array}\right)  = \left(\begin{array}{c}
\mathcal{M}^{\dagger }\chi_1\\ \mathcal{M}^{-1 } \chi_2\end{array}\right)
\ee
The induced measure on the gauge invariant fermionic Hilbert space in this case is$ e^{-2 \mathcal{S}(h)} d\check{\tilde{\Lambda}} d\tilde{\Lambda}, h = \mathcal{M}^\dagger \mathcal{M} = H^{-1}$, while 
The current for $h$ is
\be
\tilde J_r = \partial h h^{-1} = -H^{-1}\partial H\label{j}
\ee
comparing with (\ref{J}), we note that going from $A$ to $a$ has effectively exchanged the left and right currents of the WZW model.  The Maxwell term for the $a$ fields in this Hilbert space assumes the form
\begin{eqnarray}
H[a] &=& \frac{e^2}{2\pi}\int \left[-J_r\frac{\delta}{\delta J_r} + \frac{1}{(x-y)^2}\frac{\delta}{\delta  J_r(x)}\frac{\delta}{\delta  J_r(y)}\right] + \nonumber \\
&&\frac{1}{2e^2}\int b^2  
\end{eqnarray}
The negative sign in front of the first term above follows from the same in front of $\mathcal{S}$ in the measure for this theory. \\We now note that the two sets of gauge invariant variables are related as:
\be
\Lambda  = \sigma^1 \check{\tilde{\Lambda}}\label{lambdas}
\ee
Using this and (\ref{covi1}) we may transform $\Psi$ directly to $\tilde \Lambda$. The resulting transformation is
\be
\left(\begin{array}{c}\check{\tilde{\lambda_1}}\\ \check{\tilde{\lambda_2}} \end{array}\right) = (I - Im \theta+ Re \theta \gamma^5 + \cdots)\sigma^1\Psi 
\ee
The above equation tells us two things. Namely, it is the parity transform of $\Psi$ that transforms to $\check{\tilde{\Lambda}}$ and the relative change in the sign in front of the $Re \theta$ terms implies that the signs of the measures have opposite signs depending on whether $\Lambda$ or $\check{\tilde{\Lambda}}$ is used as the gauge invariant spinor corresponding to $\Psi$. To compare the corresponding Maxwell terms, we must transform the fermions to the same gauge invariant object, which we take to be $\check{\tilde{\Lambda}}$. When acting on the hilbert space spanned by $H, \check{\tilde{\Lambda}}$ and $\tilde{\Lambda}$, and after accounting for the change in the sign of the measure, the Maxwell Hamiltonian assumes the form
\begin{eqnarray}
H_{Max} &=& H[A] = \frac{e^2}{2\pi}\int \left[-J_l\frac{\delta}{\delta J_l} + \frac{1}{(x-y)^2}\frac{\delta}{\delta J_l(x)}\frac{\delta}{\delta J_l(y)}\right] \nonumber\\
&&+ \frac{1}{2e^2}\int B^2  
\end{eqnarray}
This is to be compared with $H[a]$ and the two equal each other term-by-term.
\end{document}